\documentclass{emulateapj}

\usepackage{url}

\newcommand\name[1]{{\small\sc #1}}

\slugcomment{}

\shorttitle{Random Walk Noise in Blazars}
\shortauthors{Park \& Trippe}

\begin{document}

\title{Radio Variability and Random Walk Noise Properties of Four Blazars\altaffilmark{1}}

\author{Jong-Ho Park \& Sascha Trippe\altaffilmark{2}}
\affil{Department of Physics and Astronomy, Seoul National University, Gwanak-gu, Seoul 151-742, South Korea}
\received{2013 December 18}
\accepted{2014 February 25}
\email{jhpark@astro.snu.ac.kr (J.-H.P.) , trippe@astro.snu.ac.kr (S.T.)}

\altaffiltext{1}{Based on observations obtained by the University of Michigan Radio Astronomy Observatory (UMRAO)}
\altaffiltext{2}{Corresponding author}

\begin{abstract}
\noindent We present the results of a time series analysis of the long-term radio lightcurves of four blazars: 3C 279, 3C 345, 3C 446, and BL Lacertae. We exploit the data base of the University of Michigan Radio Astronomy Observatory (UMRAO) monitoring program which provides densely sampled lightcurves spanning 32 years in time in three frequency bands located at 4.8, 8, and 14.5\,GHz. Our sources show mostly flat or inverted (spectral indices $-0.5\lesssim\alpha\lesssim0$) spectra, in agreement with optically thick emission. All lightcurves show strong variability on all time scales. Analyzing the time lags between the lightcurves from different frequency bands, we find that we can distinguish high-peaking flares and low-peaking flares in accord with the classification of \cite{Valtaoja}. The periodograms (temporal power spectra) of the observed lightcurves are consistent with random-walk powerlaw noise without any indication of (quasi-)periodic variability. The fact that all four sources studied are in agreement with being random-walk noise emitters at radio wavelengths suggests that such behavior is a general property of blazars. 
\end{abstract}


\section{Introduction \label{sect_intro}}

\noindent
The strong and complex temporal flux variability of Active Galactic Nuclei (AGN; see, e.g., \citealt{Beckmann} and references therein for a recent review) provides valuable information on the internal conditions of accretion zones and plasma outflows. Various characteristic variability patterns have been associated with a wide range of physical phenomena, from shocks in continuous (e.g., \citealt{Marscher}) or discontinuous (e.g., \citealt{spada2001}) jets to orbiting plasma ``hotspots'' (e.g., \citealt{abramowicz1991}) or plasma density waves (e.g., \citealt{kato2000}) in accretion disks. Accordingly, multiple studies have aimed at quantifying the properties of AGN variability on all time scales and throughout the electromagnetic spectrum. At radio wavelengths, variability time scales probed by observations range from tens of minutes (\citealt{Schodel}, studying the mm/radio lightcurve of M 81*; see also \citealt{Kim2013} for a discussion of the detectability of intra-day variability) to tens of years (\citealt{Hovatta2007}, in a statistical analysis of the long-term flux variability of 80 AGN). Of particular interest is the possible presence of quasi-periodic oscillations (QPO) which has been reported by several studies of blazar lightcurves (e.g. \citealt{Rani2009, Rani2010, Gupta2012}).

Fourier transform, period folding, power spectrum, and periodogram methods (cf. \citealt{priestley1981} for an exhaustive review of time series analysis) have been used extensively for quantifying the statistical properties of AGN variability and for the search for possible QPOs (e.g. \citealt{Benlloch} for X-ray, \citealt{Webb} for optical, \citealt{Fan1999} for near infrared, and \citealt{Aller} for radio observations). As already noted by \citet{Press}, power spectra of AGN lightcurves globally follow power laws $A_f{\propto}f^{-\beta}$ with $\beta>0$, corresponding to \emph{red noise};\footnote{In the context of time series analysis, the term ``noise'' refers to stochastic emission from a source of radiation, \emph{not} to measurement errors or instrumental noise.} here $A_f$ denotes the power spectral amplitude as function of sampling frequency $f$. Lightcurves composed of pure Gaussian \emph{white noise} have flat power spectra ($\beta=0$). Other important special cases are \emph{random walk noise} ($\beta = 2$) -- which is the integral of white noise -- and \emph{flicker noise} ($\beta = 1$, ``$1/f$ noise'') as intermediate case between white noise and random walk noise (see also \citealt{Park2012} for a detailed technical discussion). \citet{Lawrence1987} found that the power spectrum of the Seyfert galaxy NGC 4051 can be described by flicker noise. Red noise power spectra were also observed by \cite{Lawrence1993} who used 12 high-quality ``long look'' X-ray lightcurves of AGN.

\begin{deluxetable*}{ccccccccc}

\tablecaption{Properties of our four target blazars \label{Information}}
\tablehead{
\colhead{Object} & \colhead{RA} & \colhead{DEC} & \colhead{Type} &
\colhead{Redshift} & \colhead{$T$ [yr]} &
\colhead{$N_{4.8}$} & \colhead{$N_{8.0}$} & \colhead{$N_{14.5}$}
}
\startdata
3C 279 & 12:56:11 & $-$05:47:22 & FSRQ   & 0.536 & 32.52 & 1086 & 1337 & 1473 \\
3C 345 & 16:42:59 & +39:48:37   & FSRQ   & 0.593 & 32.54 & 1323 & 1315 & 1415 \\
3C 446 & 22:25:47 & $-$04:57:01 & BL Lac & 1.404 & 32.12 & 680  & 902  & 1088 \\
BL Lac & 22:02:43 & +42:16:40   & BL Lac & 0.069 & 32.54 & 1256 & 1315 & 1755 
\enddata

\tablecomments{J2000 coordinates, source types, and redshifts are taken from the NED. We also give the total monitoring time $T$ (in years) and the numbers $N$ of flux data points for 4.8, 8.0, and 14.5\,GHz, respectively.}

\end{deluxetable*}

A multitude of studies illustrates the difficulties of determining the statistical significance of supposed QPO signals in the power spectra of AGN lightcurves. The analysis of \cite{Benlloch} concluded that a previously reported quasi-periodic signal in X-ray lightcurves of the Seyfert galaxy Mrk 766 was actually statistically insignificant. \cite{Uttley} pointed out the importance of sampling effects leading to \emph{red-noise leaks} and \emph{aliasing}. \cite{Vaughan} gives an analytical approach to derive significance levels for peaks in red-noise power spectra. \cite{Do} demonstrated the power of Monte-Carlo techniques for deriving significance levels by comparing the red-noise power spectra of actual and simulated flux data.

The temporal flux variability of AGN can be exploited for elucidating the physical conditions within active galaxies especially at radio frequencies where monitoring observations of hundreds of targets have been conducted over several decades by various observatories. Remarkably, many studies aimed at analyzing long-term AGN radio variability do not take into account the intrinsic red-noise properties of the lightcurves. The incorrect assumption of constant (as function of $f$) significance levels in power spectra (following from the assumption of white-noise dominated lightcurves) has lead to reports of ``characteristic'' time scales which are actually not special at all (cf., e.g., \citealt{Ciaramella, Hovatta2007, Nieppola}).

Blazars, characterized by violent flux variability across the entire electromagnetic spectrum, are a subset of AGN which include BL Lacertae (BL Lac) objects and Flat Spectrum Radio Quasars (FSRQs). In accordance with the standard viewing angle unification scheme of AGN \citep{Urry}, it is commonly assumed that their observed emission is generated by synchrotron radiation -- dominating from radio to optical frequencies -- and inverse Compton emission -- dominating at frequencies higher than optical -- from relativistic plasma jets (almost) aligned with the line of sight. In order to perform a thorough study of the statistical properties of blazar emission, we analyze the lightcurves of four radio-bright blazars with strong flux variability -- 3C 279, 3C 345, 3C 446, and BL Lac -- provided by the University of Michigan Radio Astronomy Observatory (UMRAO) monitoring program of AGN. The data set comprises data spanning $\approx$32 years in time and covering three frequency bands located at 4.8\,GHz, 8.0\,GHz, and 14.5\,GHz.

\section{Target Selection and Flux Data}

\noindent
For our study we exploited the AGN monitoring data base of the 26-meter University of Michigan Radio Astronomy Observatory (UMRAO); the instrument, observations, and calibration procedures are described in detail by \cite{Aller1985}. Our analysis required the use of densely sampled high-quality lightcurves spanning several decades in time and obtained at several observing frequencies. Accordingly, we selected targets with (i) data available for all three UMRAO bands (4.8, 8, and 14.5\,GHz); (ii) continuously spanning at least 30 years in time; (iii) dense -- faster than monthly at each frequency  -- sampling over the entire monitoring time; (iv) a minimum flux (at all frequencies) of 2\,Jy; and (v) strong flux variability by factors $>$2. Our very strict selection criteria left us with a sample of four blazars: 3C 279, 3C 345, 3C 446, and BL Lac. Table~\ref{Information} provides an overview over the key properties of our targets (partially taken from the NASA/IPAC Extragalactic Database (NED)\footnote{\url{http://ned.ipac.caltech.edu/}}) and data. The median statistical error of a flux measurement was 0.09\,Jy. The lightcurves cover a time line from 1980 to 2012, slightly more than 32 years.

\section{Analysis}

\subsection{Lightcurves \label{ssect_lightcurves}}

\noindent
We selected our data for the purpose of time series analysis which can be misled by irregular sampling. In order to minimize such sampling effects, we binned our lightcurves in time such that the bin size is the time interval

\begin{equation}
\Delta t = 2T/N
\label{bin}
\end{equation}

\noindent where $T$ is the total observing time and $N$ is the number of data points; for our sources, $\Delta t$ is on the order of three weeks typically. (For the special case of regular sampling, $\Delta t$ corresponds to the inverse of the Nyquist frequency.) The final lightcurves are shown in Fig.~\ref{indices}; evidently, all four sources show strong variability on various time scales.

\subsection{Spectral indices \label{ssect_alpha}}

\begin{figure*}[!t]
\begin{center}
\includegraphics[scale=.59]{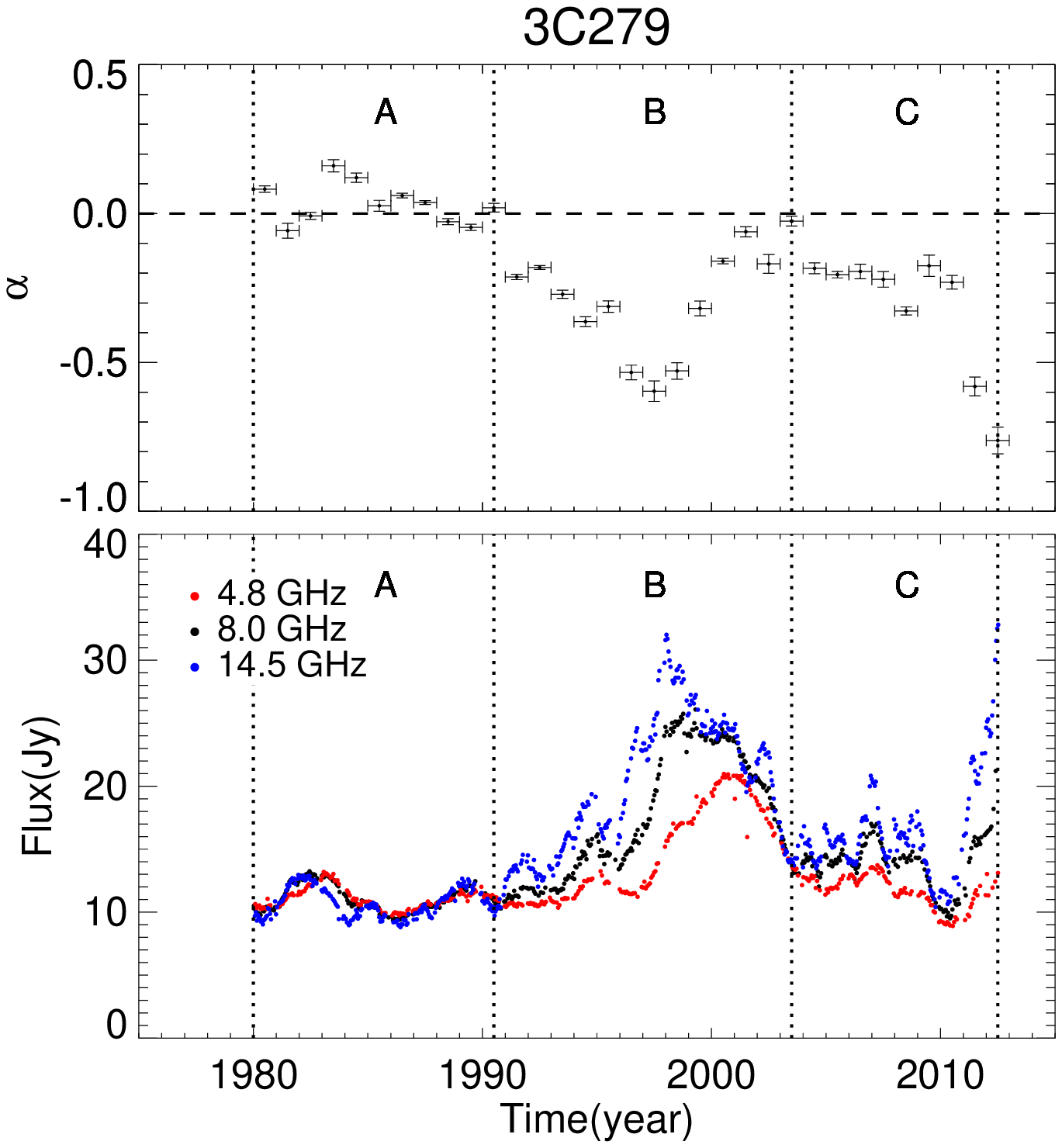}
\includegraphics[scale=.59]{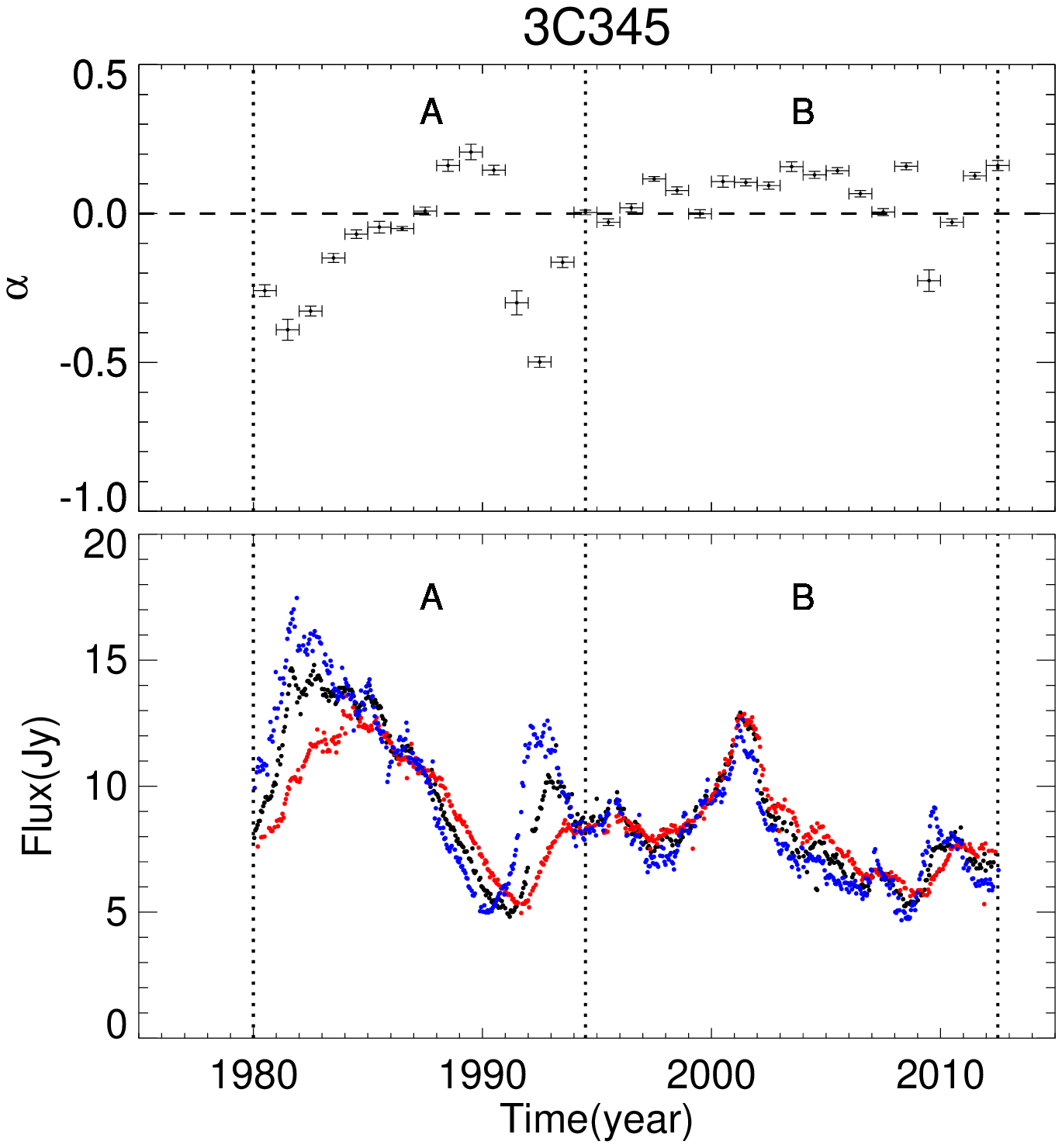} \\
\vspace{3mm}
\includegraphics[scale=.59]{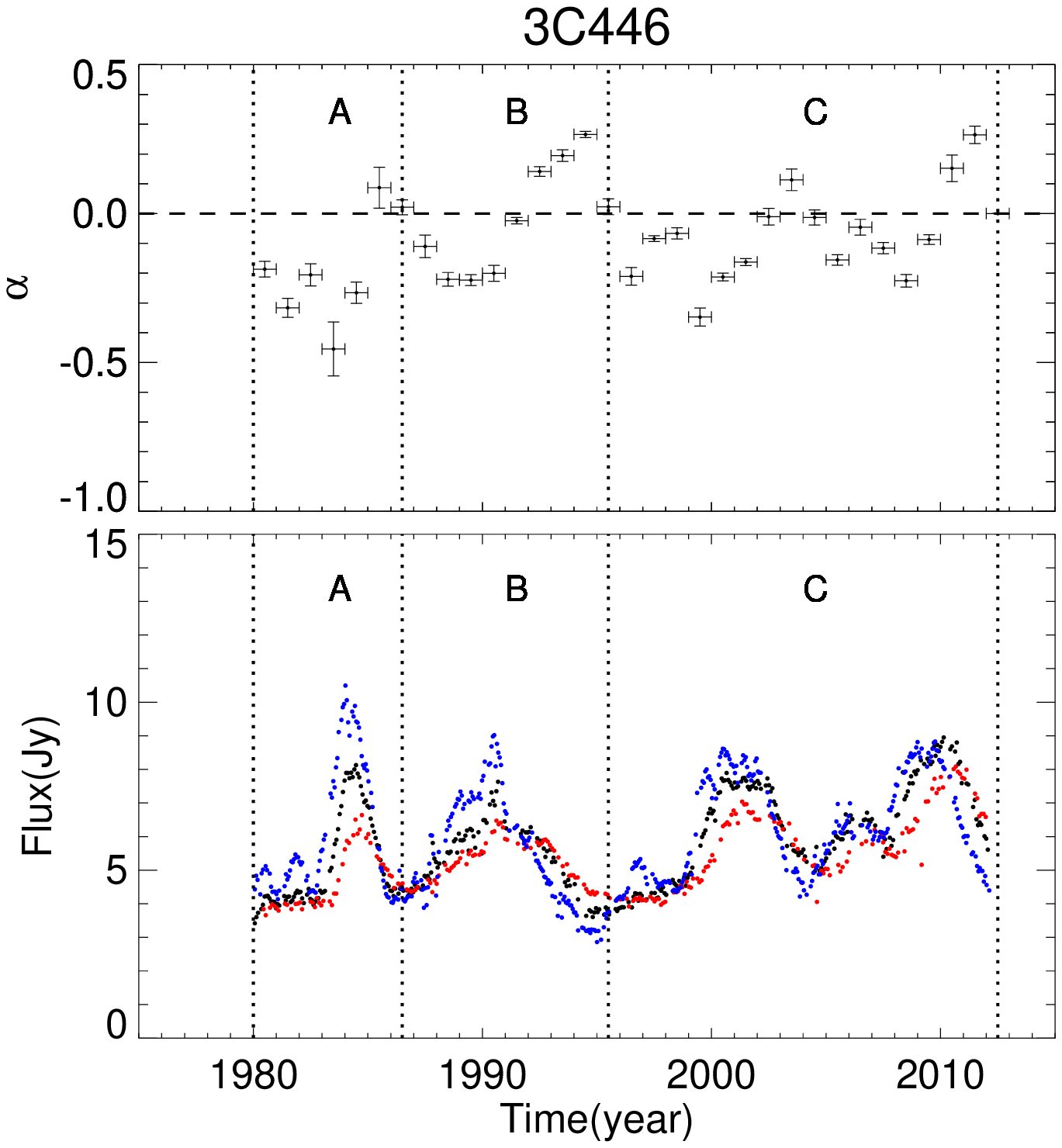}
\includegraphics[scale=.59]{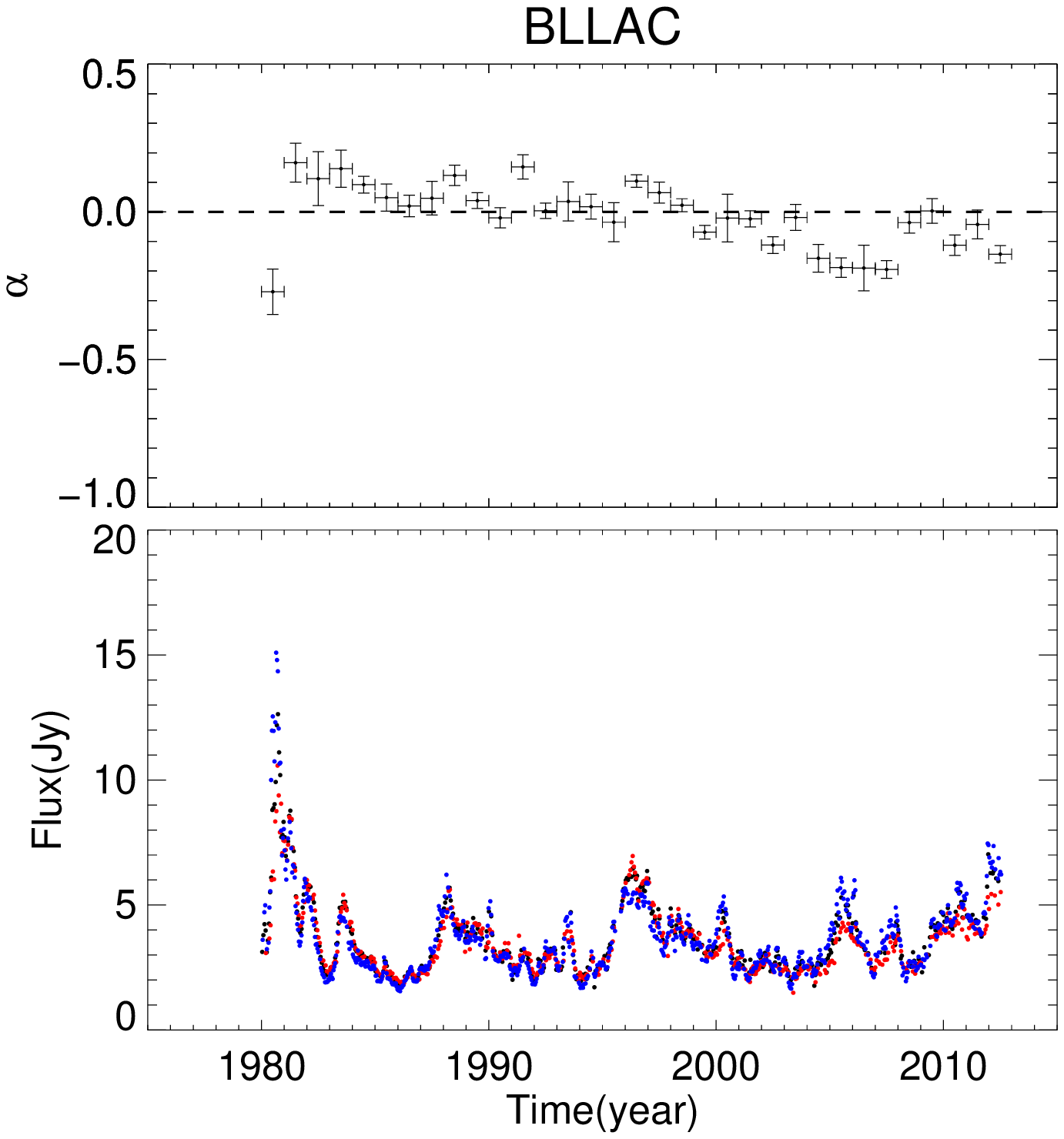}
\caption{Flux densities and spectral indices $\alpha$ as function of time for our four blazars, spanning the years 1980 to 2012. For the spectral index diagrams, error bars along the time axes denote the size of the time windows used for the calculation of $\alpha$ (1 year), error bars along the $\alpha$ axes denote the statistical $1\sigma$ errors; horizontal dashed lines show the $\alpha = 0$ lines. In the cases of 3C 279, 3C 345, and 3C 446, we divided the lightcurves into activity phases (A, B, C) with boundaries (vertical dotted lines) given by the times when $\alpha=0$ (cf. \S\,\ref{ssect_alpha}). Red, black, and blue data points indicate fluxes at frequencies 4.8, 8, and 14.5\,GHz, respectively. \label{indices}}
\end{center}
\end{figure*}

\noindent
The fast -- but not simultaneous -- sampling of our targets at three frequencies made it possible to study their spectral evolution. A combination of (i) non-simultaneous sampling and (ii) rapid intrinsic flux variability made it necessary to group our data into time windows; we eventually chose a time window of one year. Within each time window, we jointly described all data (covering all frequency bands) with a standard powerlaw model

\begin{equation}
S_{\nu} \propto {\nu}^{-\alpha}
\label{spectral_indices}
\end{equation}

\noindent where $\nu$ is the observing frequency, $S_{\nu}$ is the flux density, and $\alpha$ is the spectral index. We present the spectral index as function of time in Fig.~\ref{indices}.

\subsection{Time offsets among spectral bands \label{ssect_timelag}}

\noindent
The long overall time line and good sampling of our data made it possible to probe the data for time lags among the fluxes observed at different frequencies. We applied the discrete correlation function proposed by \cite{Edelson} to each pair of spectral bands for each source. In a first step, we computed the \emph{unbinned discrete correlation function} (UDCF) for two discrete datasets \{$a_i$\} and \{$b_j$\} with $i,j=1,2,3,...$,

\begin{equation}
{\rm{UDCF}}_{ij}(\Delta t_{ij}) = \frac{(a_i - \bar{a})(b_j - \bar{b})}{\left[(\sigma_a^2 - \delta_a^2)(\sigma_b^2 - \delta_b^2)\right]^{1/2}} ~~ .
\end{equation}

\noindent Here $\bar{a}$ and $\bar{b}$ are the averages of \{$a_i$\} and \{$b_j$\}, respectively; $\Delta t_{ij}$ denotes the difference of the observing times of the data pair ($a_i, b_j$); $\sigma_{a,b}^2$ are the variances; and $\delta_{a,b}$ denote the mean measurement errors of {$a_i$}, {$b_j$}.

The actual \emph{discrete correlation function} (DCF) for a given time offset $\tau$ results from averaging over all $N'$ ${\rm{UDCF}_{ij}}(\Delta t_{ij})$ for which $\Delta t_{ij}$ falls into a selected $\tau$ bin $\Delta \tau$ (i.e., $\tau - \Delta \tau/2 \leq \Delta t_{ij} < \tau + \Delta \tau/2$):

\begin{equation}
{\rm{DCF}}(\tau) = \frac{1}{N'}\sum^{\tau+\Delta\tau/2}_{\tau-\Delta\tau/2}{\rm{UDCF}}_{ij}(\Delta t_{ij})
\end{equation}

\noindent with ($-1$) +1 corresponding to perfect (anti-)correlation and 0 indicating the absence of any correlation. The position of the maximum of the DCF corresponds to the time offset between the lightcurves. The statistical uncertainty of DCF($\tau$) is given by the standard error of mean

\begin{equation}
\label{eq:sigmadcf}
\small
\sigma_{\rm{DCF}}(\tau) = \frac{1}{N'-1}\left[\sum^{\tau+\Delta\tau/2}_{\tau-\Delta\tau/2}\left(\rm{UDCF}_{ij} - \rm{DCF}(\tau)\right)^2\right]^{1/2} .
\end{equation}

An important effect to be considered is the interplay between (i) variations of the spectral index $\alpha$ and (ii) a time delay between the lightcurves belonging to different frequency bands: an offset between spectral bands causes variations of the observed values for $\alpha$ even if the intrinsic (i.e., corrected for the offset) spectral index is constant. Inspection of Fig.~\ref{indices} suggests that this might indeed be the case at several occasions in 3C 279, 3C 345, and 3C 446. Turning this argument around, this implies that we can divide our data into activity phases defined by the times when $\alpha=0$ (i.e. the reference spectral index for flat-spectrum AGN). Accordingly, we divided our lightcurves into two or three phases (A, B, C -- except for the case of BL Lac), and computed the DCF for each phase separately. We chose $\tau$ ranges sufficient for covering the largest time offsets expected between lightcurves, eventually adopting $\tau \pm 3.5$ years (except for the lightcurve pair 4.8/14.5\,GHz of 3C 279 where it was necessary to extend the range to $\tau \pm 5$ years). In order to preserve a good time resolution, we usually used a $\tau$ bin size $\Delta \tau = {\rm{max}}[T_1/N_1, T_2/N_2]$ for two lightcurves ``1'' and ``2''. In case of the 4.8/8\,GHz lightcurve pair of 3C 345 it was necessary to increase $\Delta\tau$ by factors up to six in order to suppress sampling artifacts.

\begin{figure*}[!t]
\begin{center}
\includegraphics[scale=.45]{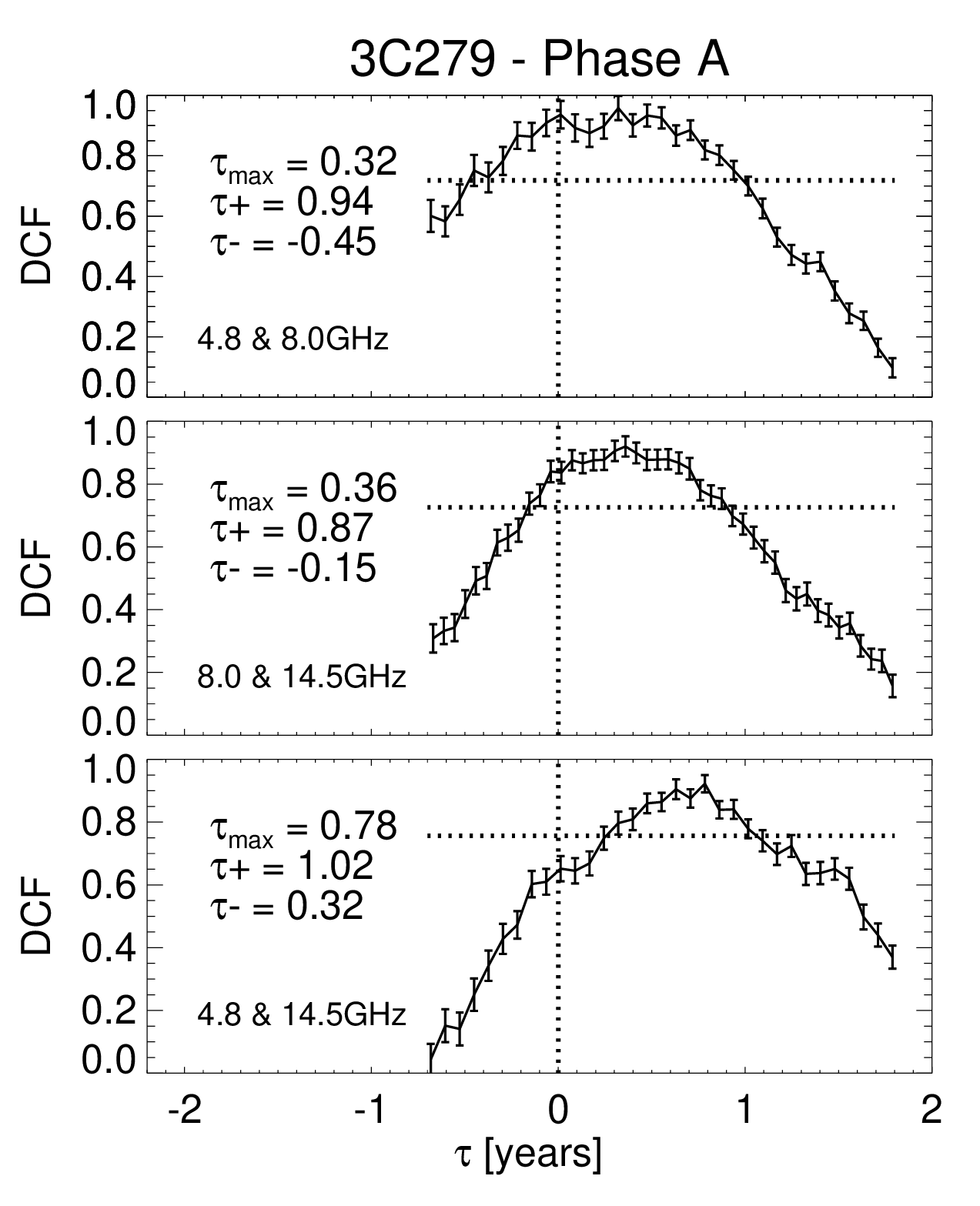}
\includegraphics[scale=.45]{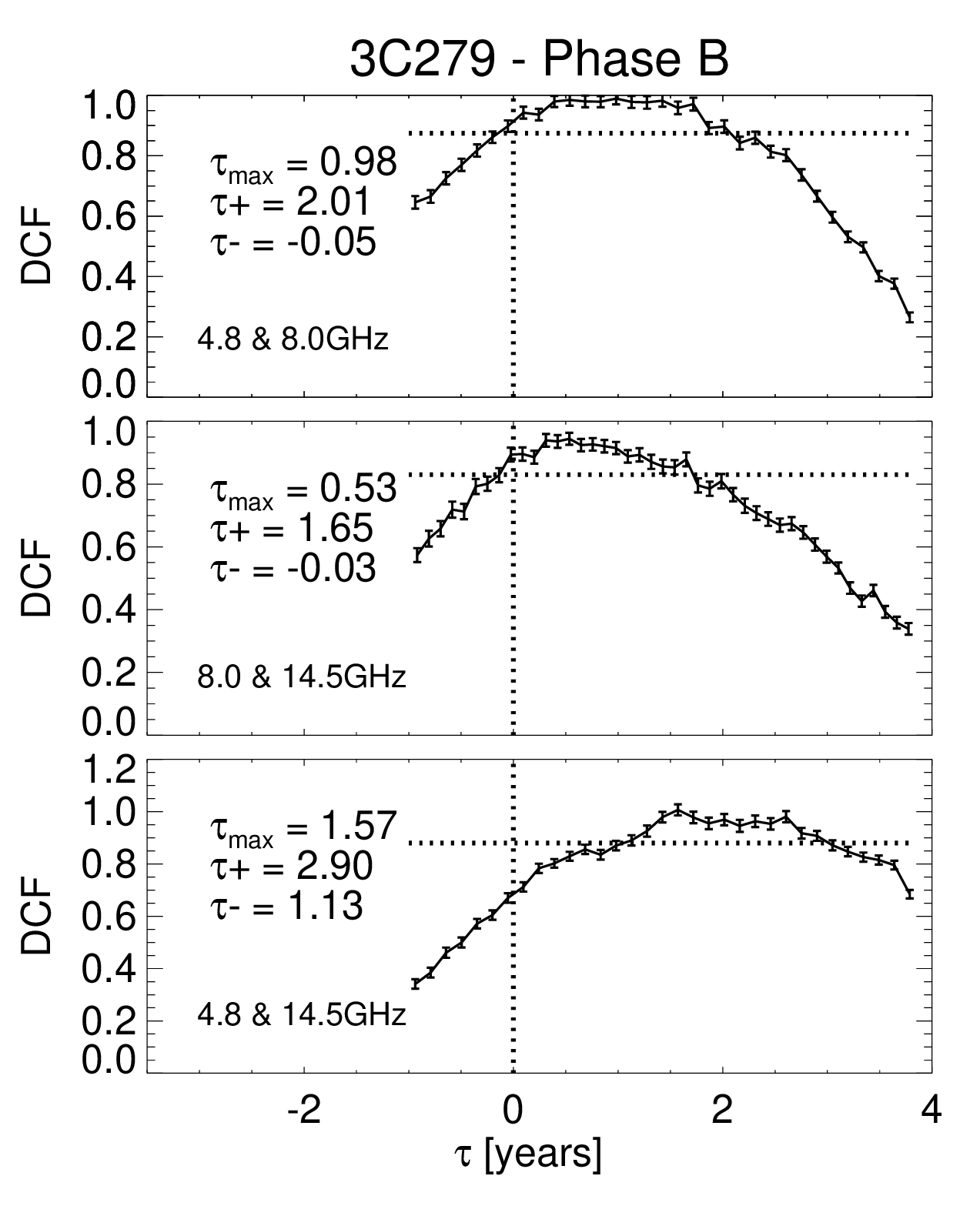}
\includegraphics[scale=.45]{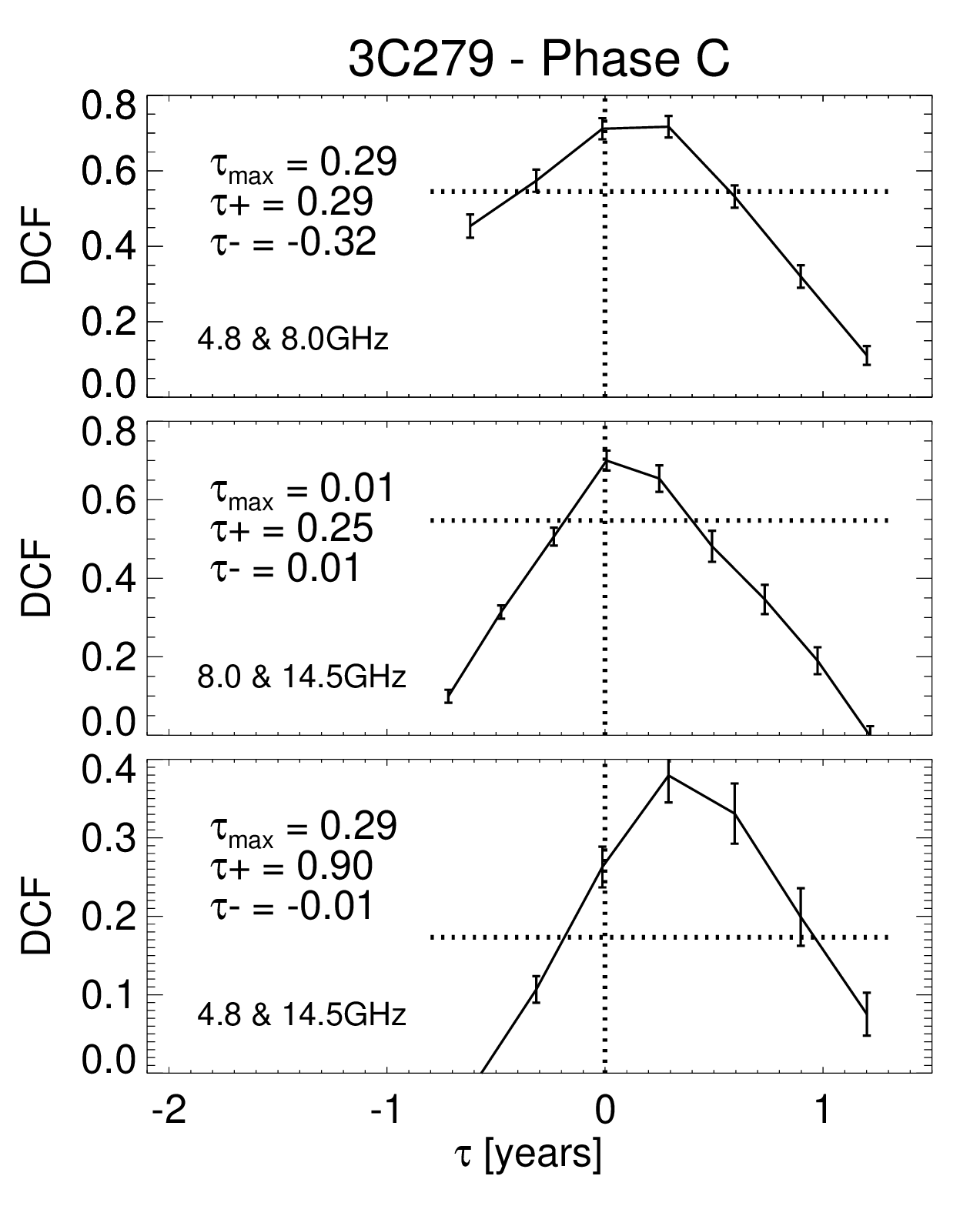} \\
\includegraphics[scale=.45]{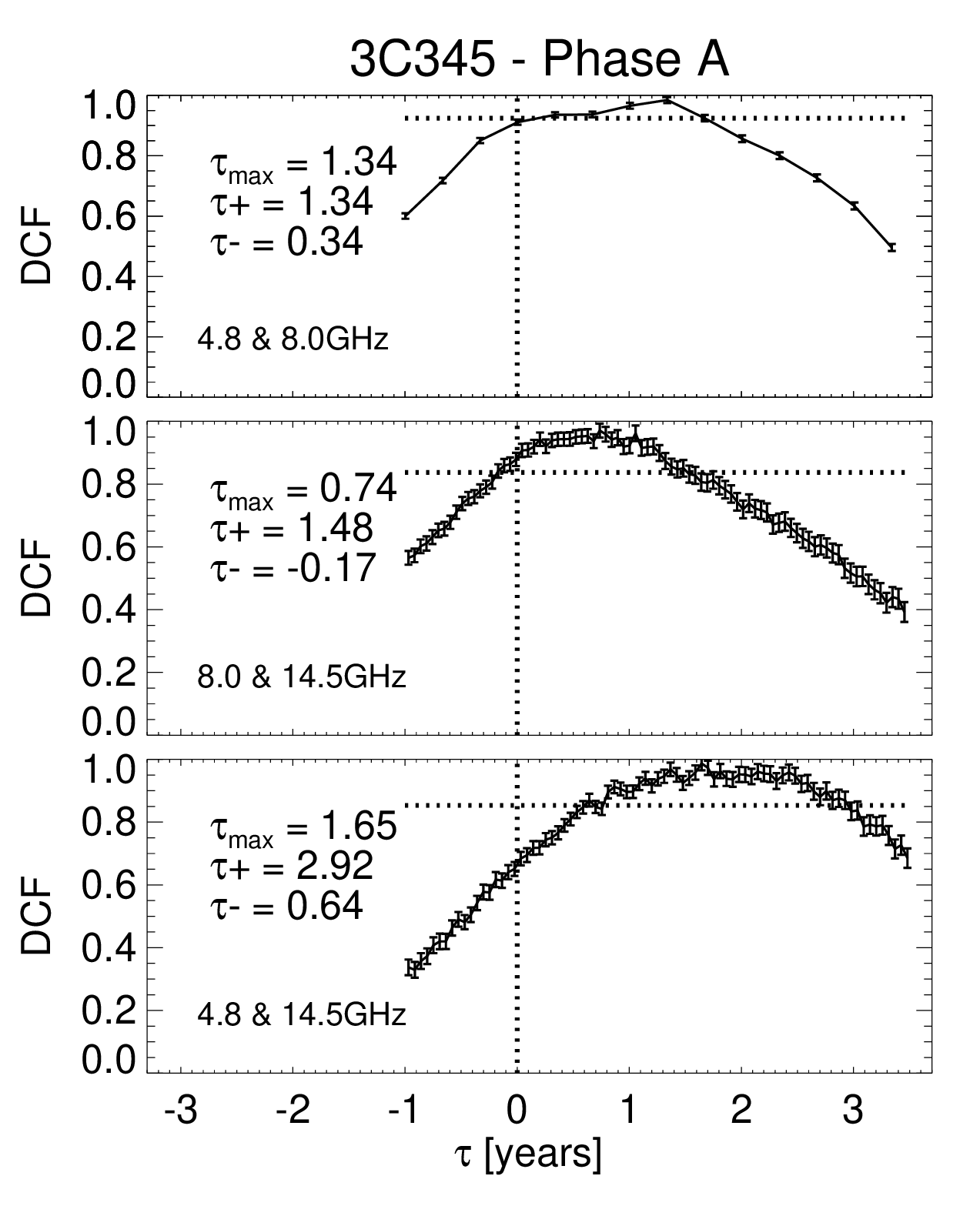}
\includegraphics[scale=.45]{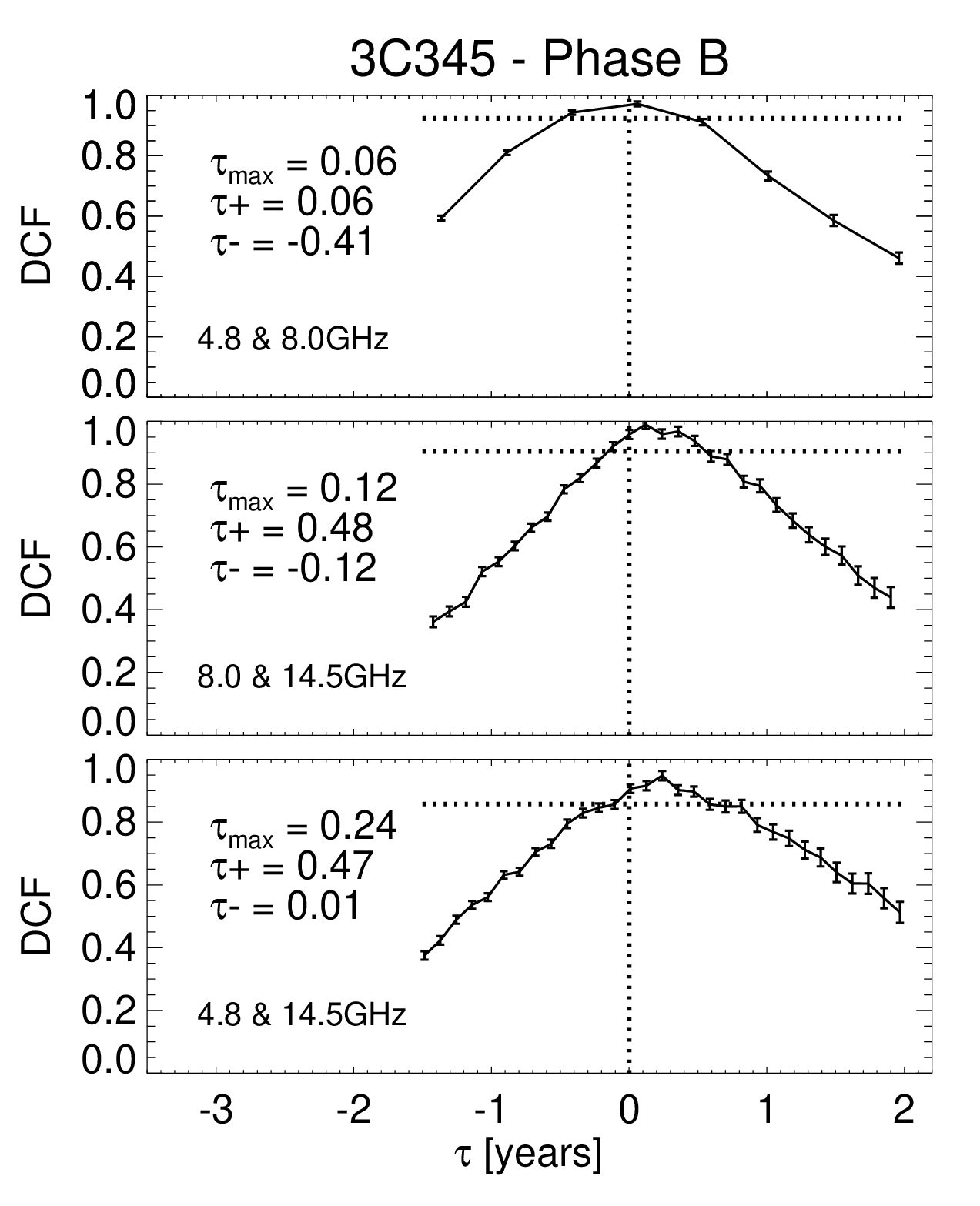}
\includegraphics[scale=.45]{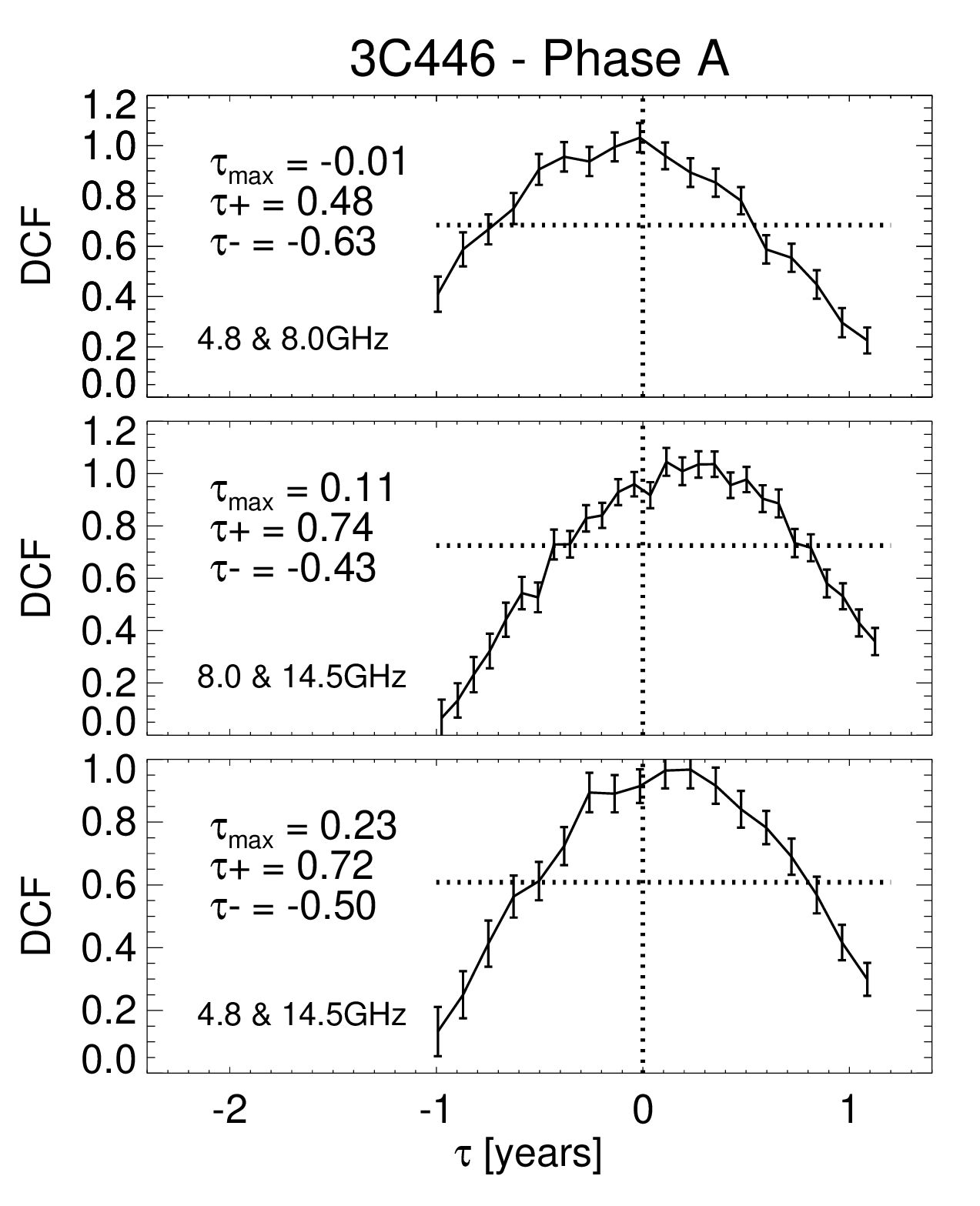} \\
\includegraphics[scale=.45]{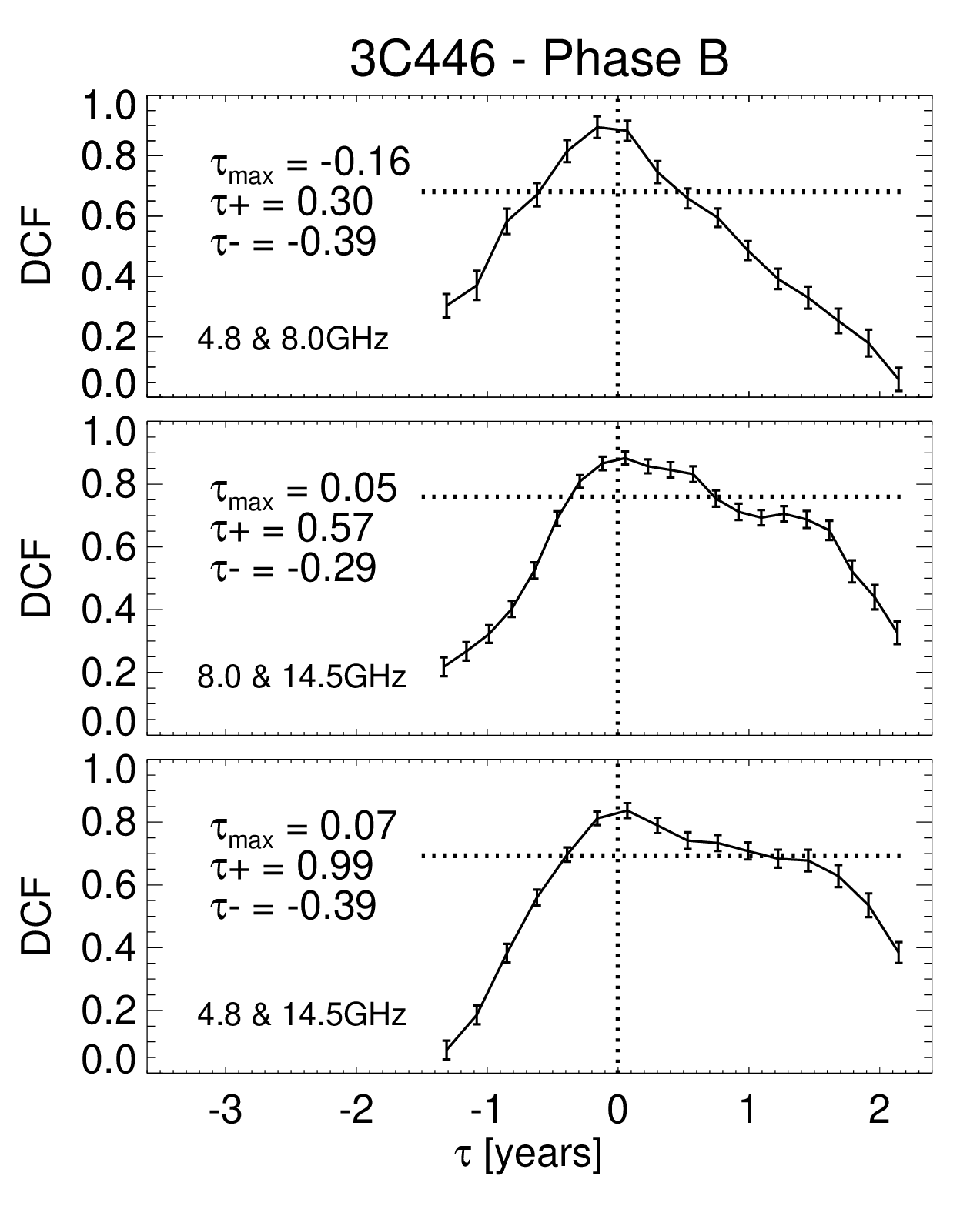}
\includegraphics[scale=.45]{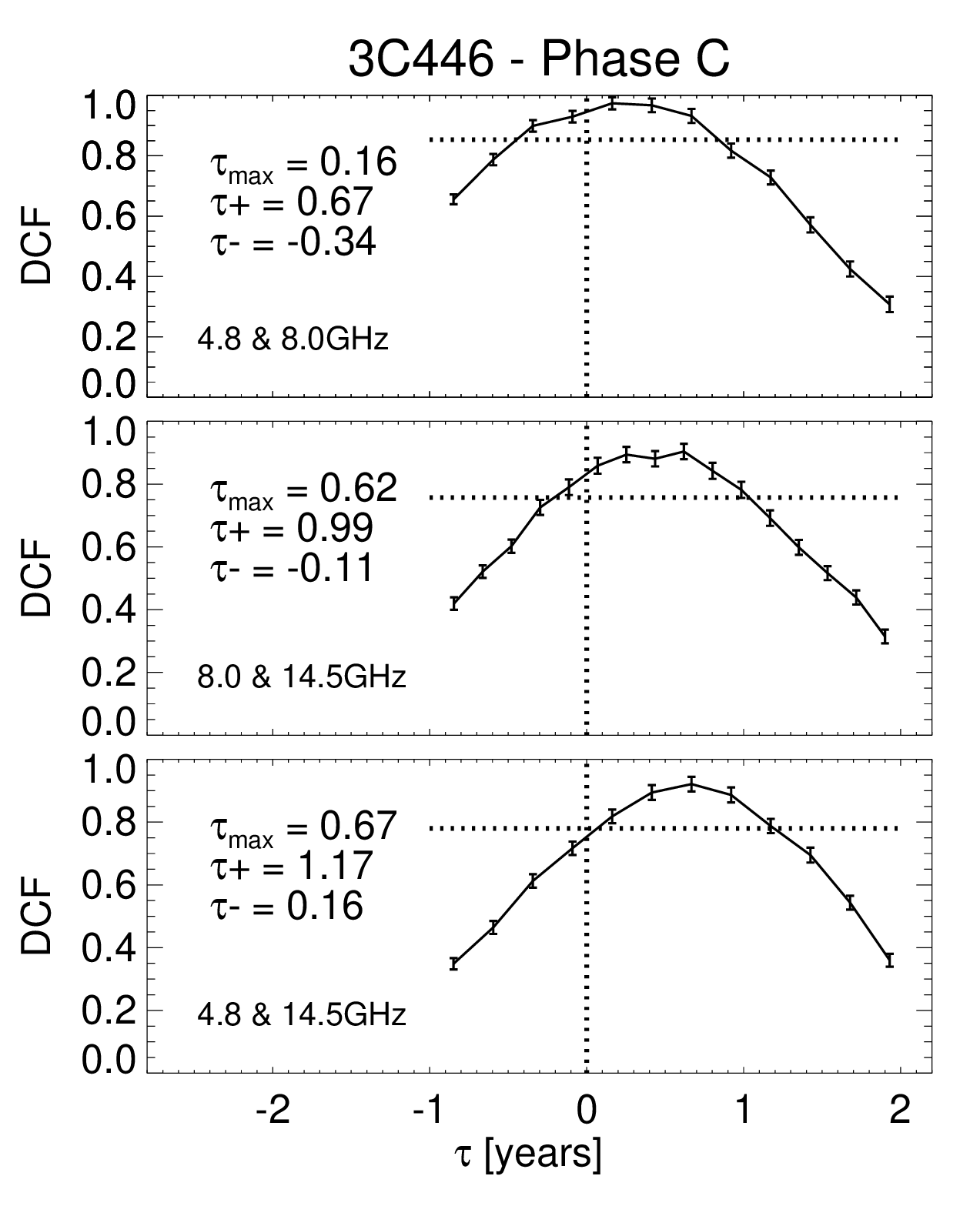}
\includegraphics[scale=.45]{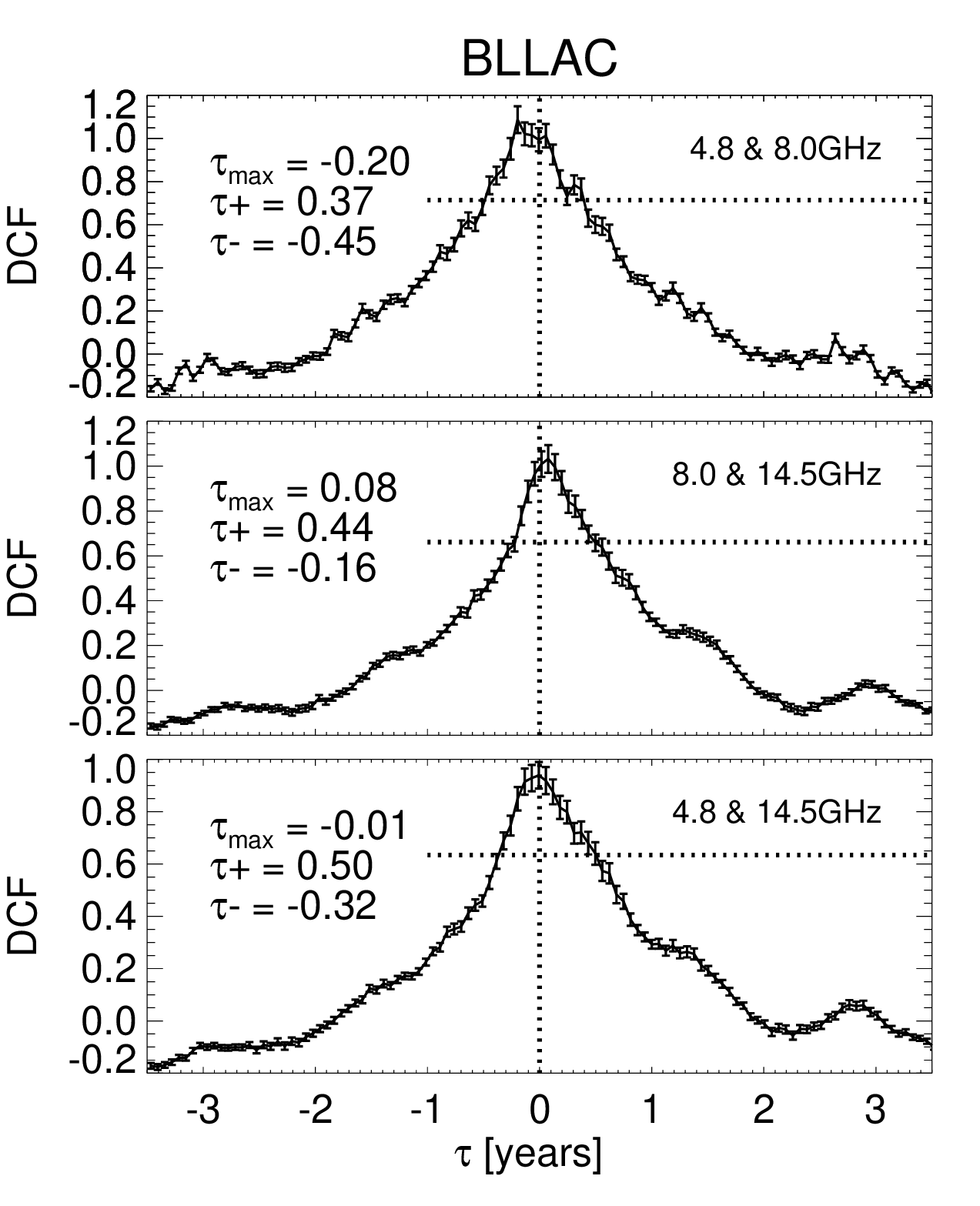}
\vspace{2mm}
\caption{Discrete correlation function (DCF) as function of time lags $\tau$ for each pair of 4.8, 8, and 14.5-GHz lightcurves of each source. The black solid lines show the DCF, error bars at each DCF point indicate the corresponding $\sigma_{\rm{DCF}}(\tau)$. The horizontal dotted lines indicate $\rm{max(DCF)} - 3\sigma_{\rm{max(DCF)}}$. The $\tau = 0$ lines are marked by vertical dotted lines. A positive (negative) time lag means that the higher frequency precedes (follows) the lower frequency. The values $\tau_{\rm max}$, $\tau_+$, and $\tau_-$ denote the time lags corresponding to the maximum of the DCF and the upper and lower boundaries of the $3\sigma$ uncertainty intervals, respectively. \label{DCF}}
\end{center}
\end{figure*}

We present the resulting DCF in Fig.~\ref{DCF}. In our convention, a positive (negative) time lag implies that the flux at the higher frequency precedes (follows) the flux at the lower frequency. As all values of a given DCF curve are computed from the same time series, adjacent DCF points are highly correlated. Depending on the $\tau$ bin size, this can cause the scatter of the DCF values to be substantially smaller than their individual errors $\sigma_{\rm{DCF}}(\tau)$ given by Eq.~\ref{eq:sigmadcf}. This implies that (i) the scatter of the DCF values is not a proxy for their uncertainties and (ii) averaging over or fitting of parametric models to multiple DCF values does not improve the accuracy of the estimate for the location of the peak of the DCF curve. With this in mind, we consider the null hypothesis ``the lightcurves are simultaneous'' as rejected if and only if DCF($\tau=0$) is located below the line defined by $\rm{max(DCF)} - 3\sigma_{\rm{max(DCF)}}$ (with $\sigma_{\rm{max(DCF)}}$ denoting the statistical $1\sigma$ error of the maximum value of the DCF).

\begin{figure*}[!t]
\begin{center}
\includegraphics[scale=.62]{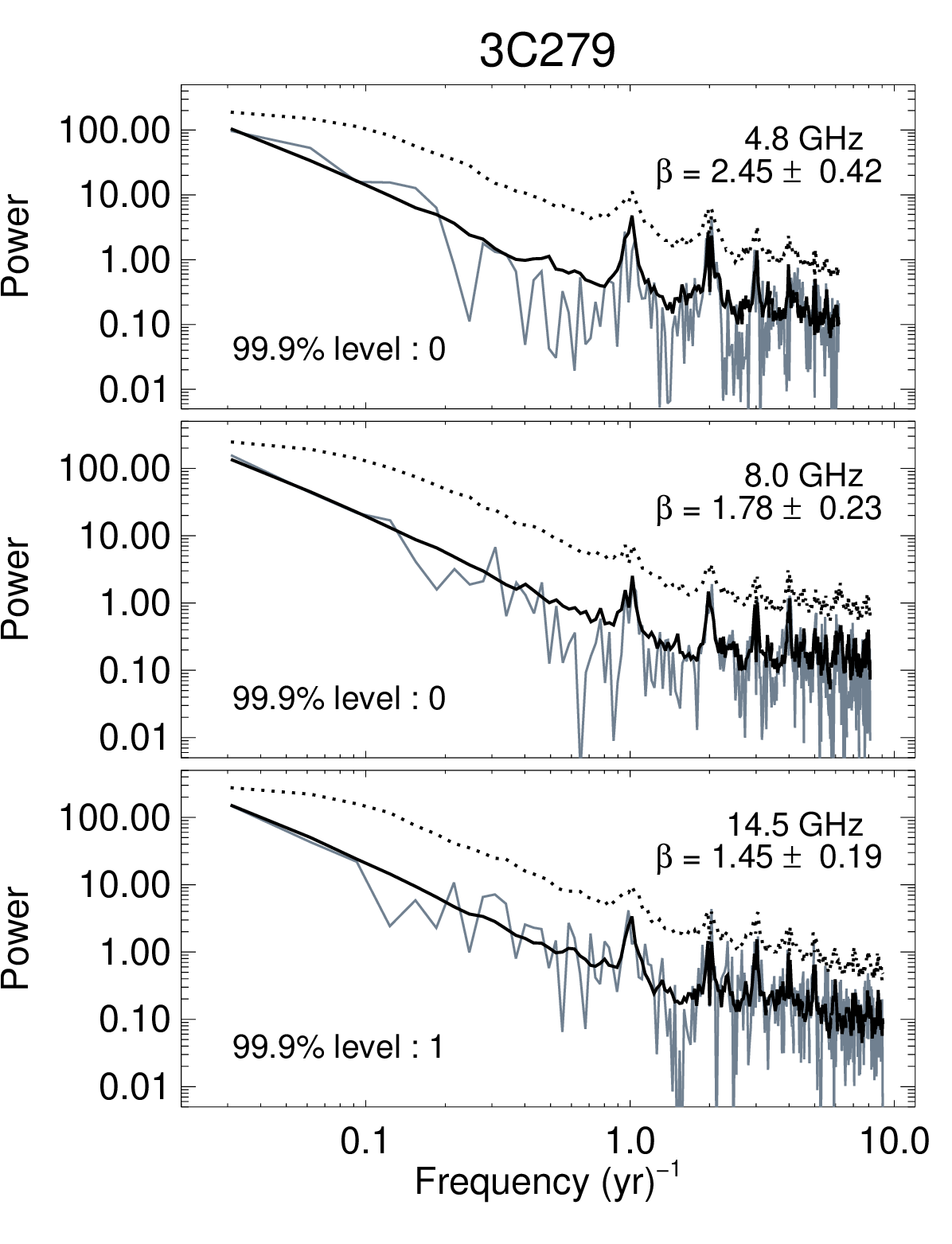}
\includegraphics[scale=.62]{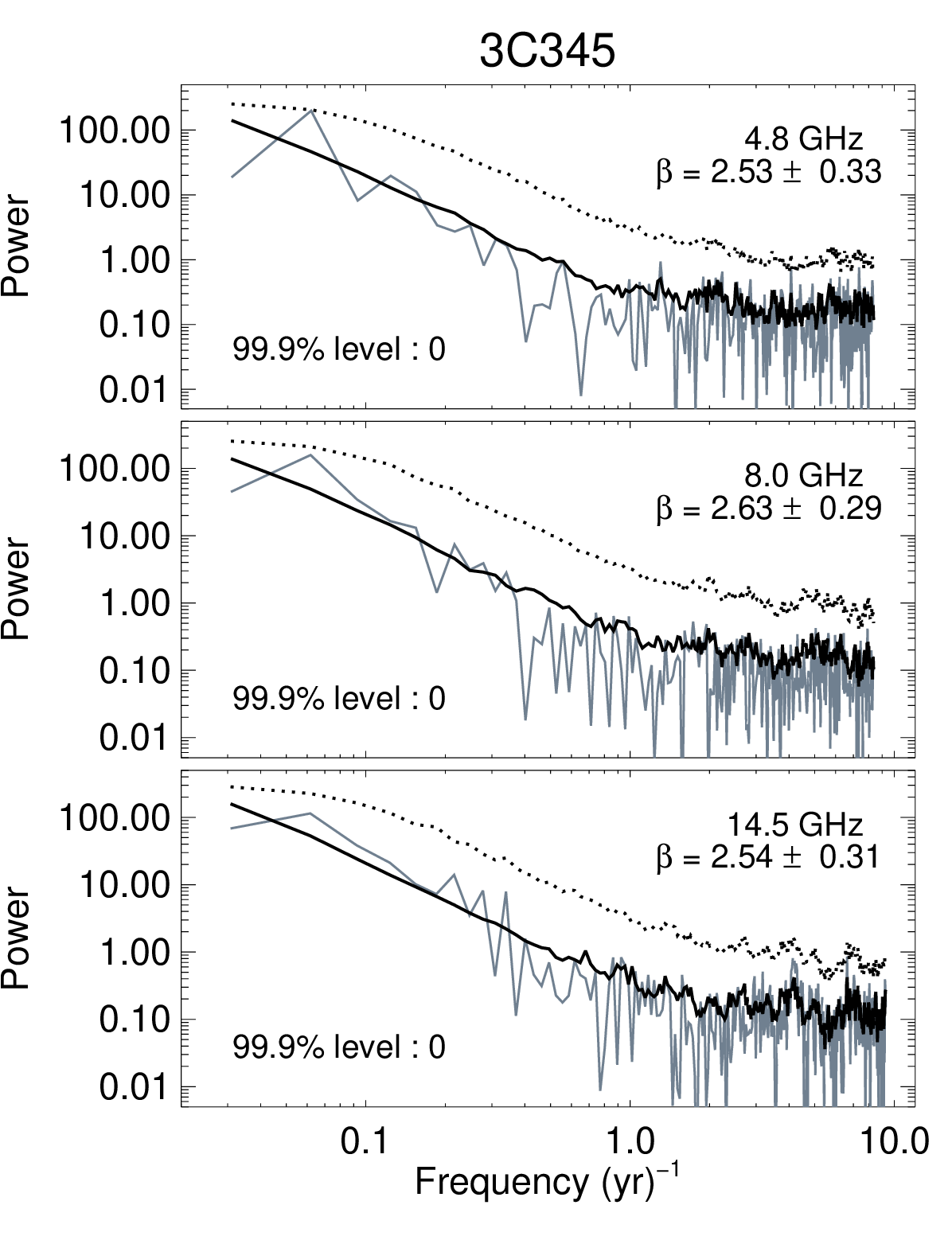} \\
\includegraphics[scale=.62]{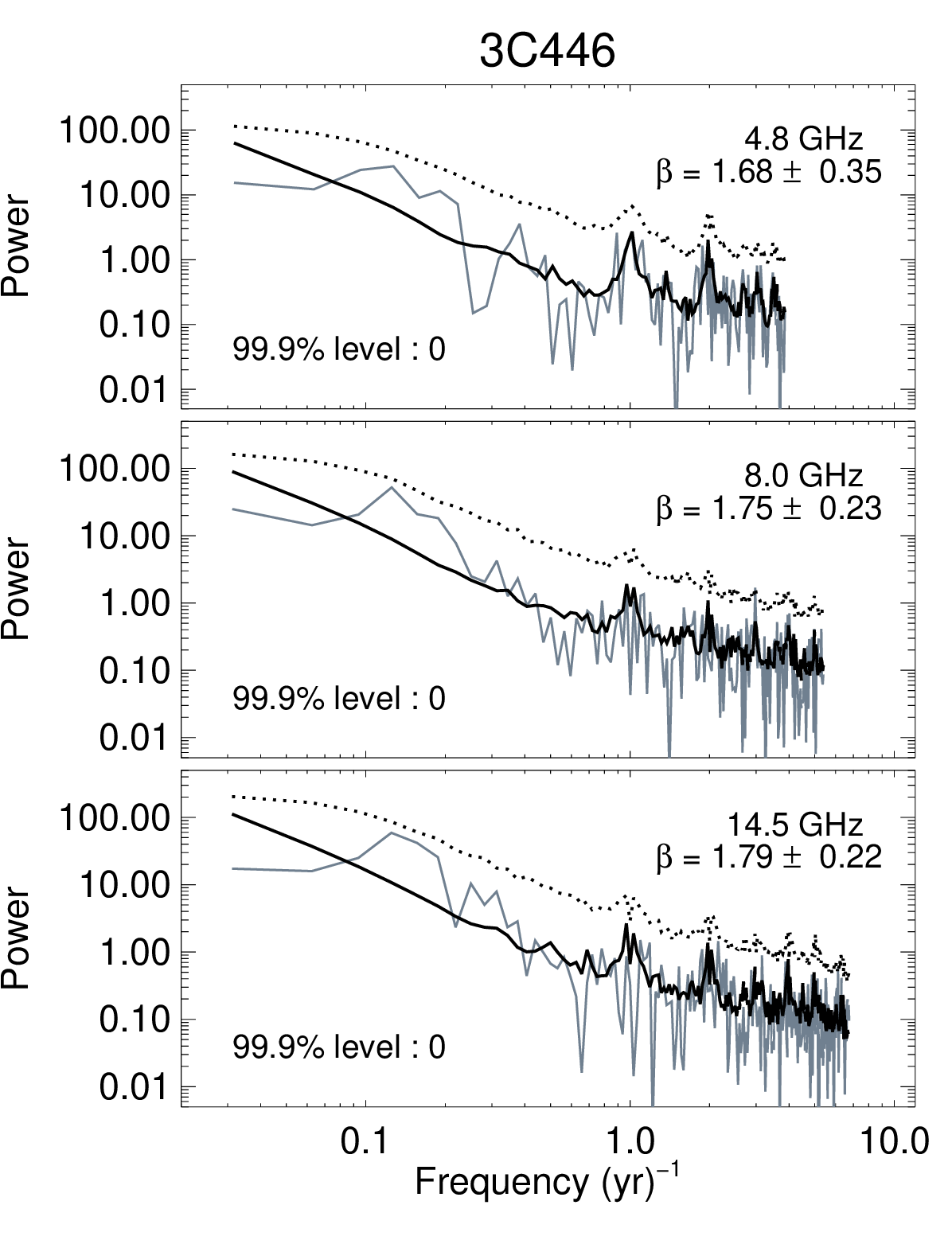}
\includegraphics[scale=.62]{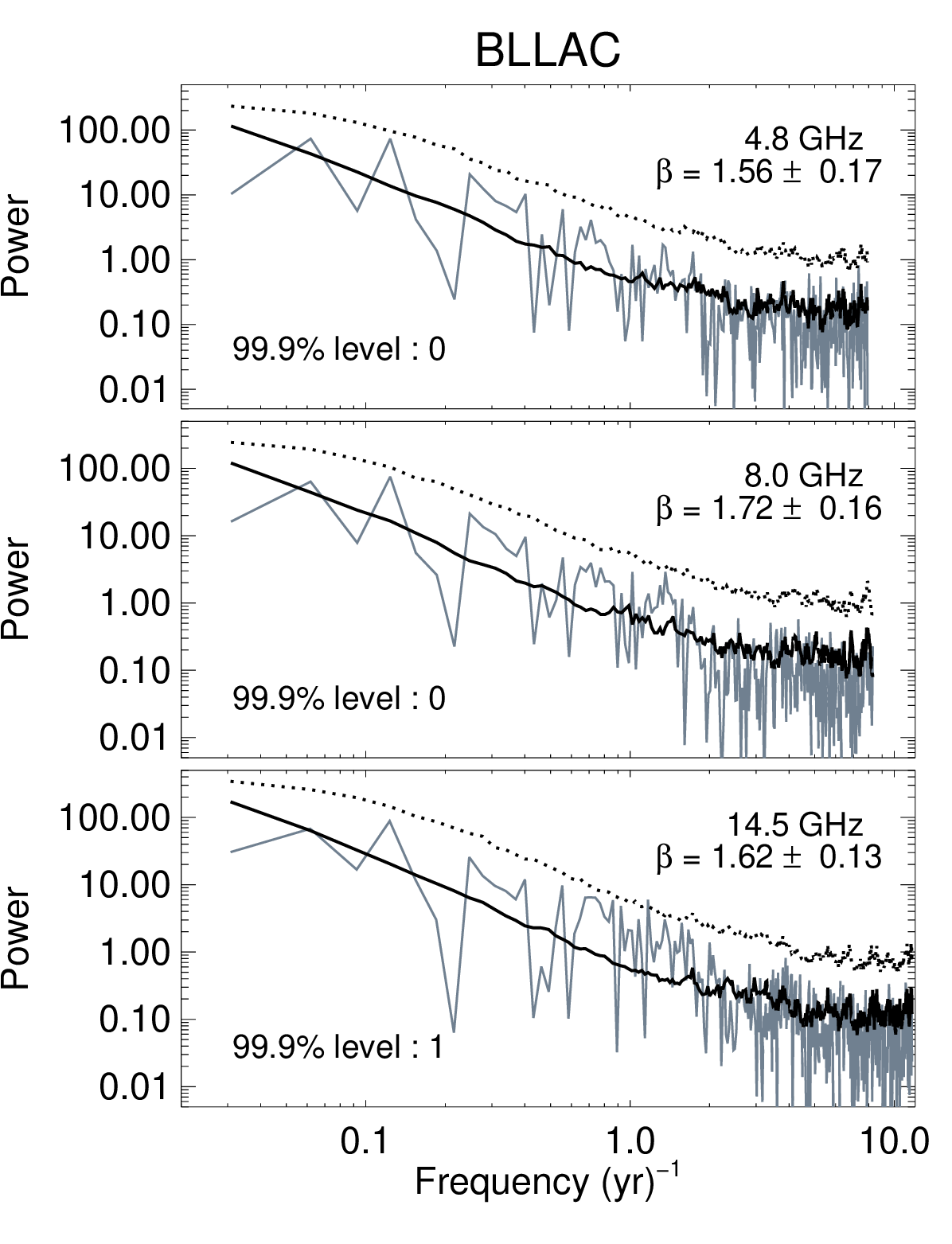}
\caption{Periodograms -- i.e. spectral power as function of sampling frequency (in units of yr$^{-1}$) -- for all 12 blazar lightcurves. Solid gray lines show the observed periodograms, solid black lines indicate the expected distributions resulting from averaging over 10\,000 simulated red-noise periodograms, and dotted black lines correspond to the 99.9\% significance levels obtained from simulations of red-noise periodograms (cf. \S\,\ref{ssect_sim}). In each diagram the value for $\beta$ found from fitting Eq.~\ref{spl} to the data (``$\beta=...$'') and the number of data points exceeding the 99.9\% significance threshold (``99.9\% level: ...'') are noted. The simulations used $\beta=2$ for 3C 279, 3C 345, and 3C 446, and $\beta=1.75$ for BL Lac. The excess values detected in the 14.5-GHz periodograms of 3C 279 (one value out of 295 frequencies probed) and BL Lac (one value out of 380) are consistent with statistical fluctuations: when taking into account the number of trials, the false alarm probabilities for these events are 26\% and 32\%, respectively. \label{Periodogram_beta}}
\end{center}
\end{figure*}

\subsection{Periodograms \label{ssect_periodo}}

\noindent
For a quantitative analysis of flux variability we employed the normalized \emph{Scargle periodogram}

\begin{equation}
\footnotesize
A_f = \frac{1}{2\sigma^2}{\left[ \frac{\Big({\sum_{i}{S_i}\,{{\cos}}\,\omega(t_i-\tau')}\Big)^2}{\sum_{i}{{{\cos}^2}\omega(t_i-\tau')}} + \frac{\Big({\sum_{i}{S_i}\,{{\sin}}\,\omega(t_i-\tau')}\Big)^2}{\sum_{i}{{{\sin}^2}\omega(t_i-\tau')}} \right]}
\end{equation}

\noindent where $\tau'$ is a time offset satisfying

\begin{equation}
\tan(2\omega\tau') = \frac{\sum_{j}\sin{2\omega t_j}}{\sum_{j}\cos{2\omega t_j}}
\end{equation}

\noindent \citep{Scargle}. Here $\omega = 2\pi f$ is the angular frequency, $A_f$ is the amplitude of the periodogram evaluated at sampling frequency $f$, $S_i$ is the $i$-th flux value, $t_i$ denotes the time when $S_i$ was obtained, and $\sigma^2$ is the variance of the data. The base frequency is $f_{\rm min}=1/T$, the sampling frequencies are $f=f_{\rm min}, 2f_{\rm min}, 3f_{\rm min}, \ldots, f_{\rm max}=N/(2T)$. Here $T$ is the total observing time and $N$ is the number of flux data points. The Scargle periodogram is preferable over standard Fourier transform methods because it can be applied to data with arbitrary sampling and has a well-understood statistical behavior \citep{priestley1981,Scargle}. We present the periodograms of our lightcurves in Fig.~\ref{Periodogram_beta}.

The power spectra of AGN lightcurves are known (cf. \S\,1) to follow red-noise powerlaws. However, when dealing with lightcurves that are sampled irregularly and show gaps, a complication arises in form of \emph{aliasing}: the power at frequencies above the Nyquist frequency is transferred to lower frequencies because variations on time scales shorter than the sampling period cannot be distinguished from variations on longer time scales. Aliasing introduces an approximately constant offset which adds to the power spectrum \citep{Uttley}. As (i) the amplitudes of power spectra tend to span several orders of magnitude and (ii) are affected by multiplicative noise \citep{Scargle,Vaughan}, periodograms have to be treated in logarithmic space. Accordingly, we assumed the functional form 

\begin{equation}
\log\left[A_f\right] = \log\left[af^{-\beta} + \delta\right]
\label{spl}
\end{equation}

\noindent for our analysis; here $a$ is a scaling factor, $\beta$ is the power-law index of the periodogram, and $\delta$ is the aliasing power. In order to estimate the powerlaw index $\beta$, we fit the model given by Eq.~\ref{spl} to each empirical periodogram; the resulting values are included in Fig.~\ref{Periodogram_beta}.

\subsection{Simulated lightcurves and significance levels \label{ssect_sim}}

\noindent
The detection of deviations from a red-noise powerlaw periodogram, especially of (quasi-)periodic signals at specific sampling frequencies, requires the establishment of reliable significance levels -- a problem that has been notoriously difficult (cf., e.g., \citealt{Vaughan}). Arguably the most straightforward ansatz is provided by Monte-Carlo simulations that compare simulated periodograms and lightcurves to actual data (e.g., \citealt{Benlloch,Do}), and this is the approach we adopted.

We simulated red-noise lightcurves using the method suggested by \cite{Timmer}. For each sampling frequency $f$, we drew two random numbers from Gaussian distributions with zero mean and unit variance for the real part and the imaginary part, respectively. We multiplied both numbers with $f^{-\beta/2}$ to generate power-law noise with slope $-\beta$. The result was an array of complex numbers which corresponded to the complex Fourier transform of the artificial lightcurve. We constructed each complex array such that the values of the Fourier transform $F(f)$ obey $F(-f_i) = F^{\ast}(f_i)$, with $^{\ast}$ denoting complex conjugation, to obtain a real valued time series. We computed artificial lightcurves $S(t)$ by taking the inverse Fourier transform of the complex arrays. 

Each artificial lightcurve consisted of 4096 data points initially. We then identified the time between two adjacent data points with the time scale defined in Eq.~\ref{bin} and omitted values from the artificial lightcurve at the locations of gaps in the observed lightcurves, thus mapping the sampling pattern of the UMRAO observations into the simulated lightcurve. Eventually, we computed a periodogram from each re-mapped artificial lightcurve. As the observed indices are $\beta\gtrsim1.5$ for all blazar lightcurves (cf. \S\,4), we used values of 1.5, 1.75, and 2 for $\beta$ in the simulations. In order to decide which value for $\beta$ to adopt for a given observed periodogram, we (i) computed 10\,000 simulated periodograms for each of the three choices of $\beta$, and (ii) compared the average of the simulated periodograms to the observed periodogram via a weighted least-squares test.

From the set of 10\,000 artificial periodograms for a given blazar lightcurve we determined, separately for each sampling frequency $f$, a 99.9\% significance level (corresponding to $3.29\sigma$ in Gaussian terms) for the periodogram derived from the UMRAO data. Our simulation procedure is based on the null hypothesis ``the observed periodogram originates from a red-noise lightcurve''. Accordingly, the spectral power of a deviation from a red-noise periodogram, especially a candidate periodic signal, needs to exceed the aforementioned significance levels in order to be potentially significant. The results of our analysis are illustrated in Fig.~\ref{Periodogram_beta}.

~~


\section{Results}

\subsection{3C 279}

\noindent
Using our criteria outlined in \S\,\ref{ssect_timelag} we divided the lightcurves of 3C 279 into three activity phases (see also Fig.~\ref{indices}). Phase A, ranging from 1980 to 1990, is characterized by relatively weak variability with fluxes ranging approximately from 10\,Jy to 14\,Jy. Flux densities in all three frequency bands are very similar for most of the time, leading to spectral indices $\alpha \approx 0$ except of the very beginning of phase A. The DCF analysis (\S\,\ref{ssect_timelag} and Fig.~\ref{DCF}) finds that the 14.5-GHz lightcurve precedes the 4.8-GHz lightcurve by $\tau=0.78^{+0.24}_{-0.46}$ years ($3\sigma$ uncertainty interval); time lags between the other lightcurve pairs are consistent with zero. Phase B, ranging from 1990 to 2003, is characterized by a strong increase in emission from $\approx$10\,Jy to $\approx$30\,Jy (at 14.5 GHz). This outburst occurs the earlier the higher the frequency; for the frequency pair 4.8/14.5 GHz, the time lag between the lightcurves is $\tau=1.57^{+1.33}_{-0.44}$ years. The observed spectral index becomes inverted, with $\alpha$ as low as about $-0.7$. Phase C, starting in 2003, is characterized by multiple flux density fluctuations in the range 10--20\,Jy for most of the time, with a strong increase -- up to 35\,Jy at 14.5\,GHz -- since 2010. The spectral index remains mildly inverted ($\alpha\approx0.2$) for most of the time but reaches $\alpha\approx-0.8$ in 2012, coinciding with the observed flux maximum at the end of the monitoring. The time lags are consistent with zero for all frequency pairs.

The periodograms (\S\,\ref{ssect_periodo}, Fig.~\ref{Periodogram_beta}) of all three lightcurves decrease toward increasing sampling frequencies -- as is characteristic for red noise -- and show a flattening at the highest sampling frequencies -- as expected in case of notable aliasing. By fitting the model given by Eq.~\ref{spl} to the spectra we find approximate powerlaw indices $\beta$ of 2.5, 1.8, and 1.5 for the 4.8-GHz, 8-GHz, and 14.5-GHz periodograms, respectively, with statistical errors between 0.2 and 0.4 ($1\sigma$ confidence intervals). Comparison of observed periodograms to the ones found from Monte-Carlo simulations (averages of 10\,000 realizations of red-noise periodograms; cf. \S\,\ref{ssect_sim}) leads to the conclusion that all three observed power spectra are consistent with being generated by random-walk noise ($\beta=2$) lightcurves. None of the observed periodograms shows statistically significant excess power with respect to a red-noise power spectrum.

\subsection{3C 345}

\noindent
We divided the lightcurves into two activity phases. Phase A, ranging from 1980 to 1994, is characterized by strong variability, with fluxes (at 14.5\,GHz) moving between 5 and 17\,Jy. This variability is also expressed in the spectral index $\alpha$ which fluctuates between $-0.4$ and 0.2. The DCF analysis finds a time lag of $\tau=1.65^{+1.27}_{-1.01}$ years ($3\sigma$ confidence intervals) between the 4.8\,GHz and 14.5\,GHz lightcurves, with smaller, marginally significant, lags between the other frequency pairs. Phase B, beginning in 1994, is characterized by strong flux variability between 5 and 13\,Jy and a mostly flat ($\alpha\approx0.2$) spectrum. Time lags between the lightcurves are (marginally) consistent with zero.

The periodograms of all three lightcurves are consistent with pure red-noise spectra. The best-fitting parametric model solutions (Eq.~\ref{spl}) show quite extreme powerlaw slopes $\beta\approx2.5$ with statistical ($1\sigma$ confidence) uncertainties of about 0.3. When comparing the data to the results of Monte-Carlo simulations, we find that all three observed periodograms are consistent with random-walk noise spectra.

\subsection{3C 446}

\noindent
During the entire observing time this source shows strong variability in both flux density -- with values ranging from 3\,Jy to 10\,Jy (at 14.5\,GHz) -- and spectral index -- with values fluctuating between $-0.5$ and 0.3. According to our criteria (\S\,\ref{ssect_timelag}) we divided the lightcurves into three phases, ranging from 1980 to 1986 (phase A), 1986 to 1995 (phase B), and from 1995 onward (phase C). Whereas in phases A and B the lightcurves of all three frequencies are consistent with being simultaneous, we find a time lag of $\tau=0.67^{+0.5}_{-0.51}$ years ($3\sigma$ confidence interval) for the pair 4.8/14.5\,GHz.

The periodograms of all three lightcurves are consistent with being pure red-noise spectra. Our parametric model fit (Eq.~\ref{spl}) finds powerlaw slopes $\beta\approx1.7$ for all three periodograms, with statistical ($1\sigma$) errors between 0.2 and 0.4. From our Monte-Carlo simulations we find that the observed periodograms are consistent with random-walk noise lightcurves ($\beta=2$).

\subsection{BL Lac}

\noindent
The lightcurve of BL Lac is characterized by rapid variability throughout the entire monitoring time of 32 years. Flux densities vary between 2\,Jy and 7\,Jy with the notable exception of a flare that reaches about 15\,Jy (at 14.5\,GHz) in 1980. The spectral index varies only slowly (except at the time of the 1980 flare) -- on time scales of years to decades -- between $\alpha\approx0.2$ and $\alpha\approx-0.2$. Accordingly, we did not attempt to identify separate activity phases -- qualitatively,  the behavior of BL Lac is actually rather uniform. The lightcurves at the three observing frequencies follow each other closely not only in amplitude but also in time: all time lags identified by the cross-correlation analysis are consistent with zero.

The periodograms of all three lightcurves are in agreement with being pure red-noise power spectra. The parametric model (Eq.~\ref{spl}) finds identical (within the $1\sigma$ errors of 0.1--0.2) $\beta\approx1.6$ for all periodograms. Comparison to the simulation results shows the best agreement with a theoretical, intrinsic powerlaw slope of $\beta=1.75$.

\section{Discussion}

\subsection{Spectral indices}

\noindent
For all sources the spectral index remains at values that are close to zero or even negative ($\alpha\lesssim0.3$). The low values of $\alpha$ imply that the emission originates from optically thick synchrotron sources, leading to approximately flat ($\alpha \approx 0$) or even inverted ($-0.5\lesssim\alpha\lesssim0$) spectra \citep{ginzburg1965,pachol1970,Kembhavi,Krolik}. This is in agreement with blazars being AGN with jets pointing (almost) toward the observer, resulting in a high column density of matter (potentially belonging to multiple individual plasma clouds) along the line of sight -- we do not find any indication for a deviation from this (expected) behavior even over a time line of three decades.

Taking a closer look at individual flux maxima (outbursts or flares of radiation), we can distinguish two types of behavior: (i) events with (a) fluxes at higher frequencies preceding the ones at lower frequencies and (b) fluxes at higher frequencies being significantly higher than the fluxes at lower frequencies (implying $\alpha<0$); and (ii) events with all lightcurves being simultaneous and of approximately equal amplitude (implying $\alpha\approx0$). A noteworthy example is provided by 3C 345, where both types of events occur within about 20 years (cf. phase A vs. phase B in Fig.~\ref{indices}). An interpretation is readily provided by the ``generalized shock model'' of \cite{Valtaoja} which is based on the assumption that outbursts of radio emission in AGN are caused by shocks propagating through jets and which distinguishes two scenarios: (i) in \emph{high-peaking flares} (``high'' with respect to the observing frequency), the maximum luminosity is reached at frequencies well above the observing frequency. This implies that the flare is decaying at the time of observation, resulting in an observational signature equivalent to the shock-in-jet model by \cite{Marscher}. This model describes shocks in AGN jets as adiabatically expanding plasmas that become optically thin at higher frequencies first, thus causing a systematic time delay between the spectral bands with the higher frequency leading with higher amplitude of flux. In case of (ii) \emph{low-peaking flares}, the maximum luminosity is reached at frequencies well below the observing frequency. Lightcurves at different frequencies are almost simultaneous and have almost identical amplitudes. Applying this framework to our sources, high-peaking flares are present in phase B and C of 3C 279, phase A of 3C 345, and the entire lightcurves of 3C 446. The other flux outbursts can be described as low-peaking flares.


\subsection{Spectral time delays}

\noindent
We examined the presence or absence of time lags between lightcurves at different frequencies for each phase of each source via discrete correlation functions. The first feature we note is the large range of time delays -- on the order of months -- permitted by the $3\sigma$ confidence limits. The different temporal evolutions of the outbursts in different spectral bands, in combination with long variability time scales on the order of years, lead to broad, asymmetric DCF curves.

As discussed already partially in the context of the spectral index analysis, we find (i) significant positive time lags, as well as (ii) phases of no or weak positive time lags between different spectral bands. Phase B of 3C 279 ($\tau \approx 0.5-1.6$ years) and phase A of 3C 345 ($\tau \approx 0.7-1.7$ years) correspond to case (i); the amplitudes reach their maxima first at 14.5 GHz, with the 8 and 4.8 GHz lightcurves trailing -- as expected for high-peaking flares \citep{Valtaoja}. The long evolution time scales of the outbursts, on the order of years, suggest physical sizes of the expanding emission regions on the order of light-years. The remaining activity phases correspond to case (ii), with time delays close to or in agreement with zero. Here we find both low- and high-peaking flares: phase C of 3C 279 and all phases of 3C 446 show very fast spectral index variability ranging from $\alpha \approx -0.8$ to $\alpha \approx 0.3$, in agreement with the behavior of high-peaking flares; BL Lac however shows simultaneous lightcurves with approximately identical amplitudes throughout the entire monitoring time of three decades -- in agreement with the behavior expected for low-peaking flares.

\subsection{Power spectra}

\noindent
The periodograms of all four blazars are in good agreement with lightcurves generated by powerlaw noise with index $\beta=2$ -- i.e., random walk noise -- and being affected by aliasing caused by irregular sampling. Furthermore, our statistical tests (\S\,\ref{ssect_sim}) show that all periodograms are consistent with being pure red-noise power spectra without significant (quasi-)periodic signals (cf. Fig.~\ref{Periodogram_beta}). The powerlaw-noise nature of their lightcurves implies that none of our target blazars shows any ``characteristic'' activity time scale.

As already outlined in \S\,1, the red-noise nature of AGN lightcurves is observationally well established (albeit this discussion is complicated by the potential presence of multiple states of emission; cf. e.g. \citealt{dodds2011,Trippe,Park2012}). Empirically, the slopes of the power spectral density of different AGN tend to scatter over a wide range of values, roughly from $\beta\approx0.5$ (e.g., \citealt{Trippe}) to $\beta\approx2$ (e.g., \citealt{Do}), with ``typical'' values $\beta\approx1$ (e.g., \citealt{Press}). Indeed, the presumed flicker-noise nature of AGN lightcurves triggered a search for an underlying physical mechanism which has lasted for more than three decades (e.g., \citealt{Press,lyubarskii1997,kelly2011}), without any clear picture emerging as yet.

Given the incoherent picture of the statistical properties of temporal AGN variability, the clarity of our results comes as a surprise: we find the lightcurves of all four blazars to be consistent with being random-walk signals ($\beta\approx2$). Within the obvious limits of low-number statistics, this suggests that random-walk noise radio lightcurves are characteristic for blazars. We note the importance of a careful treatment of data (\S\S\,\ref{ssect_lightcurves}, \ref{ssect_periodo}) as well as a careful modeling of red noise lightcurves (\S\,\ref{ssect_sim}): only the combination of good data quality, periodogram analysis, awareness of sampling effects, and Monte Carlo simulations of powerlaw noise lightcurves unveils the intrinsic random-walk noise behavior. Evidently, this raises the question if random-walk noise lightcurves could be a general feature of blazars that is frequently masked by limited data quality, irregular sampling, inappropriate modeling of power spectra, et cetera.


\section{Conclusions}

\noindent
We studied high-quality radio lightcurves of four luminous blazars -- 3C 279, 3C 345, 3C 446, and BL Lac -- spanning 32 years in time and covering the frequencies 4.8, 8, and 14.5\,GHz. We analyzed the temporal evolution of fluxes and spectral indices. Our work leads us to the following principal conclusions: 

\begin{enumerate}

\item  Our sources show mostly flat or inverted ($-0.5\lesssim\alpha\lesssim0$) spectral indices, in agreement with optically thick synchrotron emission. The lightcurves of different frequencies are either simultaneous (within errors) or shifted relative to each other such that the high-frequency emission leads the low-frequency emission by up to $\approx$1.5 years. We are able to distinguish high-peaking and low-peaking flares according to the classification of \cite{Valtaoja}.

\item  All lightcurves show variability on all time scales. Their periodograms (power spectra) are in agreement with being pure red-noise powerlaw spectra without any indication for (quasi-)periodic signals. When taking into account the sampling patterns via dedicated Monte Carlo simulations, we find that all lightcurves are consistent with being random walk noise signals with powerlaw slopes $\beta\approx2$. Given that we find this behavior in all four sources under study, this suggests that random walk noise lightcurves are a general feature of blazars.

\end{enumerate}

Our results imply that careful time series analysis of high-quality blazar lightcurves provides information on the source structure even if a target is not resolved spatially. Obviously, it will be necessary to systematically study much larger blazar samples in order to decide if the trends we have uncovered are indeed general.

\acknowledgments
\noindent
This work is based on observations obtained by the University of Michigan Radio Astronomy Observatory (UMRAO) supported by the National Science Foundation (NSF) of the U.S.A.. We are most grateful to \name{Margo F. Aller} for making the data available to us and for valuable discussion. This study made use of the NASA/IPAC Extragalactic Database (NED). We acknowledge financial support from the Korean National Research Foundation (NRF) via Basic Research Grant 2012R1A1A2041387.

~~~



\begin{thebibliography}{}

\bibitem[Abramowicz et al.(1991)]{abramowicz1991} Abramowicz, M.A., et al. 1991, A\&A, 245, 454
\bibitem[Aller et al.(1985)]{Aller1985}
Aller, H. D., Aller, M. F., Latimer, G. E., \& Hodge, P. E. 1985, \apjs, 59, 513
\bibitem[Aller et al.(2003)]{Aller}
Aller, M. F., Aller, H. D., \& Hughes, P. A. 2003, \apj, 586, 33
\bibitem[Beckmann \& Shrader(2012)]{Beckmann}
Beckmann, V., \& Shrader, C. 2012, Active Galactic Nuclei, Weinheim: Wiley-VCH
\bibitem[Benlloch et al.(2001)]{Benlloch}
Benlloch, S., Wilms, J., Edelson, R., et al. 2001, \apj, 562, L121
\bibitem[Ciaramella et al.(2004)]{Ciaramella}
Ciaramella, A., Bongardo, C., Aller, H.D., et al. 2004, \aap, 419, 485
\bibitem[Do et al.(2009)]{Do}
Do, T., Ghez, A.M., Morris, M.R., et al. 2009, \apj, 691, 1021
\bibitem[Dodds-Eden et al.(2011)]{dodds2011} Dodds-Eden, K., et al. 2011, ApJ, 728, 37
\bibitem[Edelson \& Krolik(1988)]{Edelson}
Edelson, R. A., \& Krolik, J. H. 1988, \apj, 333, 646
\bibitem[Fan(1999)]{Fan1999}
Fan, J. H. 1999, \mnras, 308, 1032
\bibitem[Fan et al.(2007)]{Fan2007}
Fan, J. H., Liu, Y., Yuan, Y. H., et al. 2007, \aap, 462, 547
\bibitem[Ginzburg \& Syrovatskii(1965)]{ginzburg1965} Ginzburg, V.L. \& Syrovatskii, S.I. 1965, ARA\&A, 3, 297
\bibitem[Gupta et al.(2012)]{Gupta2012}
Gupta, A. C., Krichbaum, T. P. Wiita, P. J. et al. 2012, \mnras, 425, 1357
\bibitem[Hovatta et al.(2007)]{Hovatta2007}
Hovatta, T., Tornikoski, M., Lainela, M., et al. 2007, \aap, 469, 899
\bibitem[Kato(2000)]{kato2000} Kato, S. 2000, PASJ, 53, 1
\bibitem[Kelly et al.(2011)]{kelly2011} Kelly, B.C., et al. 2011, ApJ, 730, 52
\bibitem[Kembhavi \& Narlikar(1999)]{Kembhavi}
Kembhavi, A. K., \& Narlikar, J. V. 1999, Quasars and Active Galactic Nuclei, Cambridge: Cambridge University Press 
\bibitem[Kim \& Trippe(2013)]{Kim2013}
Kim, J. -Y., \& Trippe, S. 2013, JKAS, 46, 65
\bibitem[Krolik(1999)]{Krolik}
Krolik, J. H. 1999, Active Galactic Nuclei: Princeton University Press 
\bibitem[Lawrence \& Papadakis(1993)]{Lawrence1993}
Lawrence, A., \& Papadakis, I. 1993, \apj, 414, 85
\bibitem[Lawrence et al.(1987)]{Lawrence1987}
Lawrence, A., Watson, M. G., Pounds, K. A., \& Elvis, M. 1987, Nature, 325, 694
\bibitem[Lyubarskii(1997)]{lyubarskii1997} Lyubarskii, Yu. E. 1997, MNRAS, 292, 679
\bibitem[Markowitz et al.(2003)]{Markowitz}
Markowitz, A., Edelson, R., Vaughan, S., et al. 2003, \apj, 593, 96
\bibitem[Marscher \& Gear(1985)]{Marscher}
Marscher, A. P. \& Gear, W. K. 1985, \apj, 298, 114
\bibitem[Montroll \& Shlesinger(1982)]{montroll1982} Montroll, E.W. \& Shlesinger, M.F. 1982, Proc. Natl. Acad. Sci. USA, 79, 3380
\bibitem[Niepolla et al.(2009)]{Nieppola}
Nieppola, E., Hovatta, T., Tornikoski, M., et al. 2009, \aj, 137, 5022
\bibitem[Pacholczyk(1970)]{pachol1970} Pacholczyk, A.G. 1970, Radio Astrophysics, San Francisco: W.H. Freeman \& Co.
\bibitem[Park \& Trippe(2012)]{Park2012}
Park, J. -H., \& Trippe, S. 2012, JKAS, 45, 147
\bibitem[Press(1978)]{Press}
Press, W. H. 1978, Comment. Astrophys., 7, 103
\bibitem[Priestley(1981)]{priestley1981}
Priestley, M.B. 1981, Spectral Analysis and Time Series, London: Elsevier
\bibitem[Rani et al.(2009)]{Rani2009}
Rani, B., Wiita, P. J., \& Gupta, A. C. 2009, \apj, 696, 2170
\bibitem[Rani et al.(2010)]{Rani2010}
Rani, B., Gupta, A. C., Joshi, U. C. et al. 2010, \apj, 719, L153
\bibitem[Scargle(1982)]{Scargle}
Scargle, J. D. 1982, \apj, 263, 835
\bibitem[Sch\"{o}del et al.(2007)]{Schodel}
Sch\"{o}del, R., Krips, M., Markoff, S., et al. 2007, \aap, 463, 551
\bibitem[Spada et al.(2001)]{spada2001} Spada, M., et al. 2001, MNRAS, 325, 1559
\bibitem[Timmer \& K\"{o}nig(1995)]{Timmer}
Timmer, J., \& K\"{o}nig, M. 1995, \apj, 300, 707
\bibitem[Trippe et al.(2011)]{Trippe}
Trippe, S., Krips, M., Pi\'{e}tu, V., et al. 2011, \aap, 533, A97
\bibitem[Urry \& Padovani(1995)]{Urry}
Urry, C. M., \& Padovani, P. 1995, \pasp, 107, 803
\bibitem[Uttley et al.(2002)]{Uttley}
Uttley, P., McHardy, I. M., \& Papadakis, I. E. 2002, \mnras, 332, 231
\bibitem[Valtaoja et al.(1992)]{Valtaoja}
Valtaoja, E., Ter$\ddot{\rm{a}}$sranta, H., Urpo, S., et al. 1992, \aap, 254, 71
\bibitem[Vaughan(2005)]{Vaughan} Vaughan, S. 2005, \aap, 431, 391
\bibitem[Webb et al.(1988)]{Webb}
Webb, J. R., Smith, A. G., Leacock, R. J., et al. 1988, \aj, 95, 374


\end{thebibliography}
\end{document}